\documentclass[manuscript]{emulateapj}

\usepackage{subfigure,epsfig,url,amsmath}

\begin{document}

\title{Very Wide Binary Stars as the Primary Source of Stellar Collisions in the Galaxy} 


\author
{Nathan A. Kaib\altaffilmark{1} \& Sean N. Raymond\altaffilmark{2,3}}

\altaffiltext{1}{Center for Interdisciplinary Exploration and Research in Astrophysics (CIERA) and Department of Physics and Astronomy, Northwestern University, 2131 Tech Drive, Evanston, Illinois 60208, USA}
\altaffiltext{2}{Universit\'e de Bordeaux, LAB, UMR 5804, BP F-33270, Foirac, France}
\altaffiltext{3}{CNRS, LAB, UMR 5804, F-33270, Floirac, France}

\begin{abstract}

We present numerical simulations modeling the orbital evolution of very wide binaries, pairs of stars separated by over $\sim$10$^3$ AU.  Due to perturbations from other passing stars  and the Milky Way's tide, the orbits of very wide binary stars occasionally become extremely eccentric, which forces close encounters between the companion stars \citep{kaib13}.  We show that this process causes a stellar collision between very wide binary companion stars once every 1000--7500 years on average in the Milky Way.  One of the main uncertainties in this collision rate is the amount of energy dissipated by dynamic tides during close (but not collisional) periastron passages.  This dissipation presents a dynamical barrier to stellar collisions and can instead transform very wide binaries into close or contact binaries.  However, for any plausible tidal dissipation model, very wide binary stars are an unrealized, and potentially the dominant, source of stellar collisions in our Galaxy.  Such collisions should occur throughout the thin disk of the Milky Way.  Stellar collisions within very wide binaries should yield a small population of single, Li-depleted, rapidly rotating massive stars.

\end{abstract}

\section{Introduction}

Until very recently, it has been assumed that physical collisions between stars can only occur within a few unique environments in our Galaxy.  First, it has been shown that stellar collisions can take place in the dense cores of globular clusters  \citep{hillsday76}.  Early works considered only collisions due to encounters between two single stars.  However, it was later shown that collision rates can be substantially higher if resonant encounters between a single and binary or between two binaries are considered  \citep{hutver83}.  In these binary encounter scenarios, an unstable triple system is temporarily formed, and the triple members can closely approach one another and physically collide before the system completely dissociates  \citep{leonard89}.  It is estimated that the Milky Way's globular cluster population yields one stellar collision every $\sim$10$^{6}$ yrs  \citep{perkrat12}.

One other area of the Milky Way in which star-star collisions are thought to occur is very near the Galactic center.  Observations of the stellar populations found very near the Galactic center have revealed that there is a notable paucity in K-giant stars (2--8 M$_{\sun}$ AGB stars)  \citep{alex99}.  One potential explanation invoked to explain this depletion is that stellar collisions are common in the innermost $\sim$0.2 pc of the Galaxy  \citep{baildav99}.  Such massive stars are large collisional targets, and it has been shown that a collision can cause the outer layers of the massive star to be expelled, leaving behind a tight (but dim) binary consisting of the other colliding star and the stripped K-giant  \citep{baildav99}.  However, a typical K-giant will require 10$^{9-10}$ yrs to undergo a collision, which is longer than their lifetimes, and thus it is not clear if the unusual stellar population near the Galactic center is a true signature of collisions.  

Very recently, two new stellar collision sources have been proposed that greatly increase the potential for stellar collisions throughout the Milky Way.  Both of these are centered around the dynamics of triple star systems.  In the first scenario, triple star systems become unstable due to the mass loss of one of the evolving member stars  \citep{perkrat12}.  In the ensuing instability, triple members can pass quite close to each other before any stars are ejected from the system.  Because the evolving star is likely an AGB star, it represents a very large collision target.  Thus, collisions between the AGB star and one of the other main sequence stars can take place often.  \citet{perkrat12} estimate that evolving triple star systems yield one such stellar collision every $\sim$10$^{4}$ yrs in the Milky Way.  Because triple stars are found throughout our galaxy, this is the first work suggesting that stellar collisions could regularly occur within Milky Way field environments.

The second stellar collision scenario also involves triple star systems. In this case, at least one of the stars is undergoing Kozai-Lidov oscillations  \citep{katzdong12, koz62, lid62}.  Collisions arise because the maximum eccentricity (or minimum periastron) of Kozai-Lidov oscillations is not fixed with time, but instead varies when higher order terms of the gravitational potential are included.  This effect can be further enhanced when mass loss due to stellar evolution is included \citep{shapthom13}.  Moreover, the periastron shift per orbit can be substantially larger than a stellar radius, which diminishes the chance of tidal circularization before collision occurs \citep{kis98}.  While \citet{katzdong12} focus on collisions between white dwarfs within triple star systems and demonstrate that they are plausible, they do not calculate the rate of main sequence stellar collisions in such systems.  Because main sequence stars are larger relative to the periastron change per orbit during Kozai-Lidov oscillations, tidal circularization may be more important for these stars, and it is not clear what the main sequence stellar collision rate is in such systems.

Both of these new collision sources are consequences of the fact that the orbital paths of the stellar members of a triple star system are not fixed with time.  Thus, stellar orbits explore phase space until they sometimes reach a collisional configuration.  However, this type of evolution is not confined to triple star systems.  Some binary star systems should also have this property.  While tight binaries (for our purposes, mean stellar separations below $\sim$10$^{3}$ AU) are unaffected by perturbations from their local galactic environment, this is not true for very wide binaries ($a >$ $\sim$10$^{3}$ AU).  Torques from the Milky Way's tide and impulses from other passing field stars continually change the periastron of very wide binary orbits \citep{oort50, heitre86, jiatre10, kaib13}.  This evolution often causes binaries to undergo brief phases of very low periastron, forcing close passages between companion stars.  Recently it was shown that the low-$q$ excursions of very wide binary star systems can regularly trigger the disruption of planetary systems orbiting the member stars  \citep{kaib13}.  Disruptions of solar system-like planetary systems occur when the periastron falls below $\sim$100 AU, which is quite common.  However in much more extreme instances, the periastron of very wide binaries could approach values as small as a stellar radius, and collisions between stellar companions could occur.  

In this work, we determine the rate that very wide binary star systems are driven to periastron values comparable to a stellar radius.  Based on this, we predict the rate of stellar collisions within very wide binary star systems.  Our work is organized into the following sections: Section 2 describes the numerical methods we employ to model stellar collisions within very wide binaries.  Section 3 presents the results of these numerical simulations and predicts the stellar collision rate in the Milky Way due to very wide binaries.  Section 4 outlines an analytical understanding of the collision process and an estimate of the collision probability of very wide binary stars.  In Section 5 we discuss a potential observational signature of collisions within very wide binaries, and finally, we summarize our work in Section 6.

\section{Numerical Methods}

\subsection{Initial Conditions}

We use a simple symplectic algorithm to integrate the orbital evolution of $5\times10^{5}$ different very wide binary star systems for 10 Gyrs.  To increase computing efficiency, these systems are split into batches of 1000 binaries.  In these 1000-binary batches, each binary is assumed to have the same primary star mass, whose mass is randomly drawn from a mass function of the present day solar neighbourhood \citep{reid02}.  Once the primary mass is chosen, the secondary mass for each of the 1000 binaries is drawn separately from a uniform mass distribution between 0.1 M$_{\sun}$ and the primary mass \citep{rag10}.  With the stellar masses of the binaries decided, we next assign orbits to each of the binaries.  Semimajor axes are randomly selected from a log-uniform distribution between $10^3$ and $3\times10^4$ AU \citep{op24, pov07}.  The rest of the orbital elements are drawn from an isotropic distribution.

\subsection{External Perturbations}

Our binaries are perturbed by the tide of the Milky Way's disk as well as impulses from other passing field stars.  For the Milky Way's tide, we employ a model containing both radial and vertical terms \citep{lev01}.  This tidal model implicitly assumes that the systems being simulated (very wide binaries in our case) are on fixed circular orbits about the Galactic center.  In this tidal model, the vertical term is approximately an order of magnitude stronger than the radial term at the Sun's galactocentric distance ($\sim$8 kpc), and the strength of this term is set by the local density of matter in the disk.  In the Sun's case, the gas disk and the stellar disk of the Milky Way make roughly equal contributions to the local galactic density of matter \citep{justjah10}.  However, the scale height of the gas disk is approximately three times less than the stellar disk \citep{justjah10}.  As a result, most stars in the Milky Way lie further away from the Milky Way's midplane than the Sun, and the local galactic matter surrounding them will actually be dominated by the stellar component.  Because of this, we assume the density of matter is dominated by stars when calculating the strength of the Galactic tide in our binary simulations.  At $r=8$ kpc, we assume the total local galactic density is 0.04 M$_{\sun}$/pc$^3$.  

The other external perturbation we must model is the effect of field star passages.  Rather than modelling these with direct integrations, we employ the impulse approximation \citep{rick76}.  For our binary systems, this is a perfectly acceptable approximation since the orbital periods of even our $a=10^3$ AU systems are much longer than the encounter timescales of the vast majority of relevant and/or plausible stellar passages.  To generate individual stellar passages, we employ a method similar to that used in studies of the Oort Cloud \citep{rick08}.  Stellar masses are selected from the observed local mass function \citep{reid02}, and encounter velocities are selected from the observed mass-dependent stellar velocity dispersions seen in Hipparcos data \citep{garc01}.  The rate of encounters for each stellar mass category is set by the local dispersion and spatial density observed in Hipparcos for that particular mass range.

Rather than generating and applying a unique set of stellar encounters to each one of our $5\times10^5$ binaries, we instead reuse the same set of stellar encounters 10$^3$ times.  New sets of stellar encounters are only generated for each batch of 10$^3$ binaries mentioned in the previous section.  Thus, our simulations only use 500 different sets of stellar encounters.  

In addition to modelling the orbital evolution of binaries in the Sun's region of the Milky Way disk, we also simulate binaries at four other galactocentric distances: two that are closer to the Galactic center and two further away.  Each galactic position is separated by one disk scale length ($\sim$2.6 kpc \citep{jur08}) from the others.  The only thing that changes in these simulations is the strength of the Galactic tide and the encounter rate of passing field stars.  These are modulated by decreasing or increasing the local density according to $\rho=\rho_{0}{\rm e}^{r/r_L}$, where $r_{L}$ is the Milky Way disk's scale length and $\rho_{0}$ is the approximate density of stellar matter near the Sun's galactocentric distance ($\sim$0.04 M$_{\sun}$/pc$^3$).  When modeling external perturbations in other regions of the Milky Way's disk we do not vary the stellar dispersion.  As shown in Sections 3.1 and 4.1, as long as stellar encounters remain in the impulsive regime, the collision probability within wide binaries is not strongly dependent on field star encounter velocities.  The physical reason for this is that although an increase in dispersion increases the total number of stellar encounters, they become weaker on average, since their encounter timescales are shortened by the increased encounter velocities. 

\subsection{Dynamic Tides}

When stars pass very near one another, we must account for the dissipation of orbital energy via dynamic tides.  Our understanding of this process remains very uncertain.  Consequently, we choose to employ a very simple treatment of this dissipation for each periastron passage \citep{pt77,leeos86}.  In this model, the secondary star deposits energy into oscillations of various spherical harmonics of index $l$ within the primary:

\begin{equation}
\Delta E_{l} = \left(\frac{Gm_{1}^{2}}{R_1}\right)\left(\frac{m_{2}}{m_{1}}\right)^2\left(\frac{R_1}{q}\right)^{2l+2}T_{l}(\eta)
\end{equation}
where $m$ is stellar mass, $R$ is stellar radius, $q$ is periastron, and the subscripts refer to the primary and secondary star.  $T_{l}$ is a dimensionless parameter that depends on the stellar masses and radii as well as $q$.  The values of $T_{l}$ are of the same order for different $l$ as long as the periastron and stellar parameters are fixed.  Thus, as long as $q/R_{1} \gtrsim 3$ there is less than a $\sim$1\% error if we neglect the higher order terms and only calculate the $l=2$ (quadrupole) and $l=3$ (octopole) terms.  In our simulations, $T_{2}$ and $T_{3}$ are calculated during each periastron passage and the resulting energy change is applied instantaneously at periastron passage.  (The energy that the primary star deposits into the secondary is also calculated and applied of course.)  Our dissipation calculations are incorrect for $q/R_{1} < 3$, but as Figure 3 shows, none of these systems ever actually collide since tidal circularization begins near $q/R_{1} \simeq 5$.  

The values of the $T_{l}$ terms also depend on the assumed internal structure of the star.  The fiducial model we employ assumes all stars to be $n=3$ polytropes.  Because $n=\frac{3}{2}$ polytropes are more applicable to low-mass stars, we also rerun our simulations assuming these polytropes instead \citep{leeos86}.  As we will see in our results, our collision rates are nearly unaltered between these two models.  The reason is that the modest variation in $T_{l}$ values for different polytropes is overwhelmed by the steep drop of the quadrupole and octopole terms with $q$.  We refer readers to the original texts describing these models for the details in calculating $T_{l}$ terms \citep{pt77,leeos86}.  It should be noted that $n=3$ polytrope model in \citet{pt77} contains arithmetic errors corrected in \citet{leeos86}.  

\subsection{Numerical Algorithm}

Our binaries are integrated using a simple symplectic scheme for 10 Gyrs, roughly the age of the Milky Way thin disk.  In this kick-drift-kick scheme, the secondary is drifted along a Keplerian orbit about the primary for a timestep $\tau$ (typically $\tau=550$ yrs).  In between these drift timesteps, the acceleration due to the Galactic tide is applied as well as any impulses from passing field stars.  As stated previously, we run batches of 10$^3$ binary systems, all sharing the same primary star and external perturbations.  This scheme of shared primaries and stellar encounters greatly enhances our computing efficiency, with only a marginal loss in statistics since we still run 100 different batches of 10$^3$ binaries for each galactic position.

\subsection{Flagging Collisions}

In our simulations we define a stellar collision at any point that a binary passes through periastron, and the periastron value is below the combined physical radii of the primary and secondary stars.  Although stellar radii vary with mass, metallicity and age, there is an approximate linear one-to-one relationship between stellar mass and radius between 0.1 and 1.0 M$_{\sun}$ \citep{bayoro06}.  Based on this, we assume that our simulated stars have radii according to the following empirical relationship:

\begin{equation}
R_{*}(R_{\sun}) = 0.0324 + 0.9343m_{*}(M_{\sun}) + 0.0374m_{*}^{2}(M_{\sun}).
\end{equation}
These radii are used in calculating collision rates as well as calculating the tidal dissipation described in previous sections.

\section{Simulation Results}

To measure the probability of stellar collisions within very wide binaries, we simulate the evolution of $5\times10^5$ very wide binary systems under the influence of their local galactic environments.  Although collisions are rare, we find that they do occur.  Figure 1A shows one example of a simulated collision involving a 0.40 M$_{\sun}$ primary and a 0.18 M$_{\sun}$ secondary.  This binary begins with a semimajor axis ($a$) of $10^4$ AU and a periastron ($q$) near 2000 AU.  Perturbations from the Galactic tide and passing field stars dramatically alter this binary's periastron.  After 2.2 Gyrs, it exceeds 3000 AU, but by $t=4.85$ Gyrs periastron is driven to an extremely low value of $\sim$$10^{-3}$ AU.  This is less than the combined stellar radii of the binary members, and the two stars collide during the subsequent periastron passage.  

\begin{figure}
\centering
\includegraphics[scale=.45]{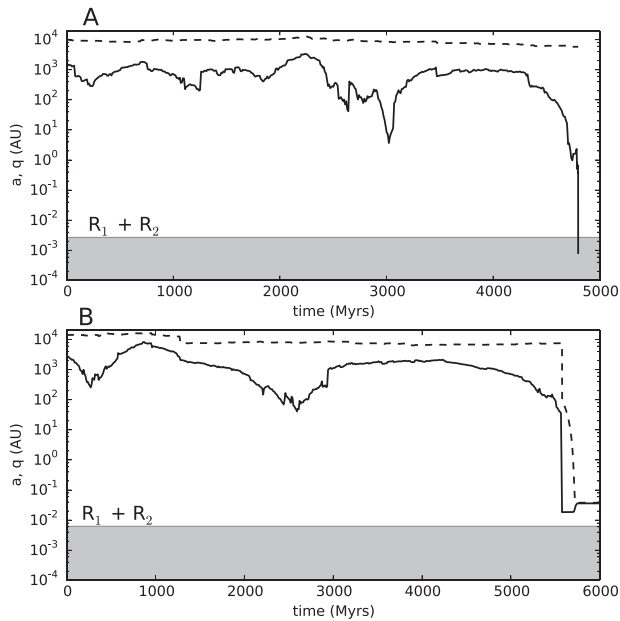}
\caption{The periastron ({\it solid}) and semimajor axis ({\it dashed}) of two different very wide binaries vs. time.  The shaded region marks the companion stars' combined radii.  Radius values are inferred from the observed stellar mass-radius relation \citep{bayoro06}.  {\bf A:} A stellar collision occurs in this system at $t=4.85$ Gyrs. {\bf B:} Dissipation from dynamic tides prevent these stars from colliding.  For illustrative purposes this example is followed until complete tidal circularization using an equilibrium tidal dissipation model below $a<100$ AU \citep{bol12} (see Section 3.1.1 for additional tide details).  In real systems the transition between dynamic and equilibrium tidal dissipation will be much smoother than that shown here.}\label{fig:1}
\end{figure}

Orbits causing star-star collisions occupy a tiny fraction of possible binary orbital parameter space.  Consequently, the rate of stellar collisions among a sample of very wide binaries depends on how quickly their orbits explore this space, and this depends on the strength of the Galactic tide and passing star impulses.  The strength of these perturbations is governed by the local density of matter in the Galaxy, which is a function of radial and vertical position in the Milky Way's disk.  Thus, to estimate the overall rate very wide binary companions collide, we model these systems in several different galactic environments.  Our modelling results are shown in Figure 2.  Here the evolution of $10^5$ very wide binary systems is modeled for 10 Gyrs at 5 different local galactic densities, corresponding to midplane disk positions between $\sim$3--13 kpc from the Galactic center.  Figure 2 shows that the galactocentric distance (local density) modulates collision rates among very wide binaries by nearly an order of magnitude.  From these results, we predict wide binaries in our Galaxy have a median probability of 1 in $\sim$500 of yielding a stellar collision.  Assuming a 5\% very wide binary fraction for stars, we predict one collision every $\sim$$10^3$ years in the Milky Way from these results.

\begin{figure}
\centering
\includegraphics[scale=.44]{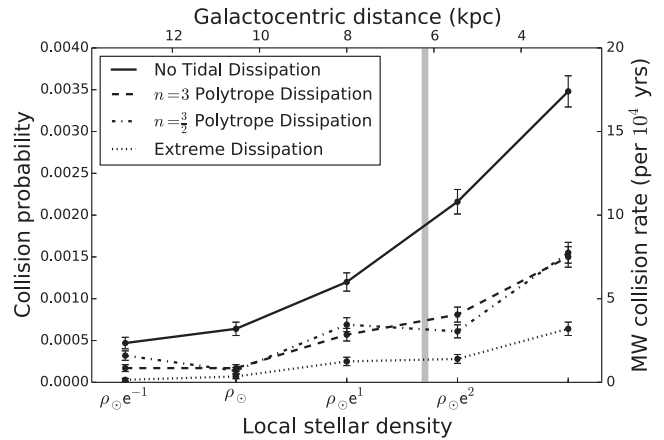}
\caption{The collision probability of simulated binary stars (with 1-$\sigma$ error bars) is plotted against their local stellar densities ($\rho_{\sun} \simeq 0.04$ M$_{\sun}$/pc$^3$).  Simulation batches with four different types of tidal dissipation are shown: no dissipation ({\it solid}), our fiducial dissipation that models stars as $n=3$ polytropes \citep{pt77,leeos86} ({\it dashed}), dissipation modelling stars as $n=\frac{3}{2}$ polytropes \citep{leeos86} ({\it dash-dot}), and our fiducial dissipation scaled up by 10 orders of magnitude ({\it dotted}).  The plotted stellar densities are converted to a galactocentric distance assuming R$_{\sun}$ = 8 kpc, a Milky Way thin disk scale length of 2.6 kpc \citep{jur08}, and stellar positions near the midplane.  Collision probabilities are converted to collision rates assuming the thin disk contains $10^{11}$ stars, and 5\% are very wide binaries \citep{dhit10, lepbon07}.  The shaded band marks the approximate median environmental density of stars in the Milky Way's thin disk.}\label{fig:2}
\end{figure}

This first batch of simulations did not include tidal interactions between binary members.  When stars come within a few stellar radii of each other they exert enormous mutual dynamic tidal forces that dissipate energy and shrink the binary orbit \citep{pt77, hut81}.  This smaller binary orbit can render galactic perturbations ineffective at causing collisions, since these perturbations only affect very weakly bound (widely separated) binaries.  Tides therefore can decrease the stellar collision rate and instead produce close binaries from wide binaries.  An example of this evolution is found in Figure 1B, which shows a simulation including a simple dynamic tidal dissipation model \citep{pt77,leeos86}.  Here, a binary has its periastron driven to $\sim$0.02 AU, roughly three times the binary members' combined radii.  At such small periastron, tidal forces rapidly dissipate orbital energy, decreasing the binary's semimajor axis from 8000 AU to $\sim$100 AU in just seven periastron passages.  Galactic perturbations cannot alter the periastron of such a tight binary, and the orbit continues to shrink at roughly fixed $q$ without colliding.  This raises the surprising possibility that very wide binaries ultimately produce a significant fraction of close or contact binaries.

Thus, there exists a ``tidal barrier'' to collisions.  If companion stars undergo a close enough encounter to activate tidal dissipation, a collision is avoided.  To assess the impact of dynamic tides on collision rates we rerun the dissipationless simulations in Figure 2 employing the dynamic tide model from  Figure 1B.  Although more sophisticated dynamic tide models exist \citep[e.g.,][]{chernov13,ivan13}, this simple treatment broadly gauges the importance of tidal dissipation.  As Figure 2 shows, tidal dissipation significantly decreases very wide binary collision rates.  For our median environmental density, collision probability decreases from 1 in $\sim$500 to 1 in $\sim$1300, or one collision every $\sim$2500 years in the Milky Way.  

Tidal dissipation becomes important when the energy dissipated in a single periastron passage approaches the binary orbital energy.  This is demonstrated in Figure 3A.  Here we show the cumulative distribution of periastron passages during one orbit {\it before} a collision occurs.  In our dissipationless simulations, the majority of colliding binary members first pass within a few stellar radii of one another before colliding on the next orbit.  However, the cumulative distribution from our tidally dissipative simulations shows that most of these collisions should never actually occur.  For a collision to occur in our tidally dissipative simulations, the companion stars must never pass within 5 radii of each other before the collision-causing periastron passage.  Figure 3B shows that this is in fact the periastron at which the energy dissipated by tides in a single periastron passage is comparable to very wide binary orbital energies.  Thus, the edge of the ``tidal barrier'' sits near 5 stellar radii.

\begin{figure}
\centering
\includegraphics[scale=.45]{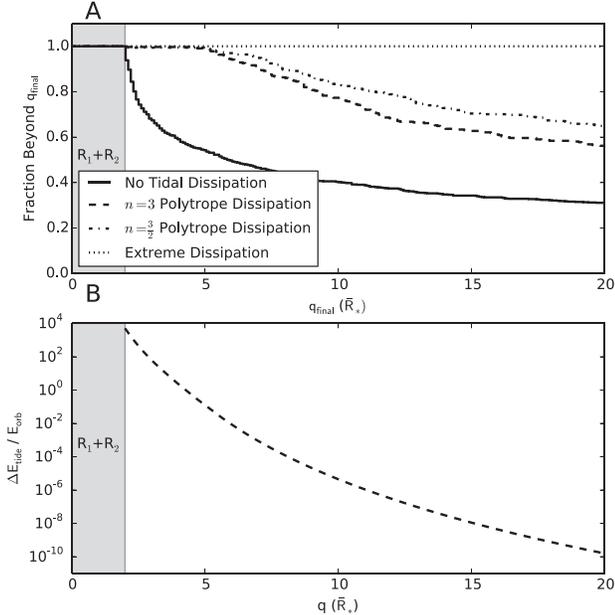}
\caption{{\bf A:} Cumulative distributions of periastron values that colliding wide binaries have one orbital period {\it before} collision.  Periastron is calculated in terms of the mean stellar radius for each system.  Distributions are shown for four different tidal dissipations:  no dissipation ({\it solid}), our fiducial dissipation that models stars as $n=3$ polytropes \citep{pt77,leeos86} ({\it dashed}), dissipation modelling stars as $n=\frac{3}{2}$ polytropes \citep{leeos86} ({\it dash-dot}), and our fiducial dissipation scaled up by 10 orders of magnitude ({\it dotted}).  The shaded region marks periastron values resulting in stellar collisions.  {\bf B:} For our fiducial tidal dissipation \citep{pt77,leeos86}, the ratio of the energy dissipated by tides during a single periastron passage to the orbital energy of a ``typical'' very wide binary is plotted against periastron.  Our ``typical'' binary orbital energy assumes both stars have masses of 0.4 M$_{\sun}$ and $a=10^4$ AU. }\label{fig:3}
\end{figure}

Our understanding of dynamic tidal dissipation remains extremely uncertain, and we must consider how much further collision rates are suppressed if our tidal model underestimates dissipation.  Although dynamic tides prevent $\sim$60\% of the collisions in our dissipationless simulations, Figure 3A shows that many collisions come from a very long tail of binaries that never undergo close encounters before colliding.  Because tidal dissipation falls off rapidly with encounter distance, suppressing a substantial portion of remaining collisions requires extremely powerful tidal dissipation.  For instance, Figure 3A shows that decreasing the collision rate by another factor of $\sim$2 requires a tidal barrier beyond at least 20 stellar radii.  According to Figure 3B, this would imply our fiducial tidal dissipation model is incorrect by at least 10 orders of magnitude!  Thus, the collision rates in our work will not drastically change if our tidal model is remotely close to reality.

To demonstrate this, we rerun the dissipationless simulations two additional times using alternative tidal models.  Our fiducial tidal model approximated stars as $n=3$ polytropes, a reasonable assumption for stellar masses $\gtrsim$ 1 M$_{\sun}$.  In our first rerun, we instead employ a tidal model approximating stars as $n=\frac{3}{2}$ polytropes \citep{leeos86}, which is well-suited for lower stellar masses.  As shown in Figure 2, this change makes almost no difference in collision rates.  Further, Figure 3A shows that the position of the tidal barrier is nearly unaltered, since both models' dissipation falls so steeply with distance.  In our final rerun, we employ a wildly dissipative model that is simply our fiducial model scaled up by 10 orders of magnitude.  Even with this extreme dissipation, many collisions from the long tail of Figure 3A still occur, and we predict very wide binary companions collide once every $\sim$7500 years in the Milky Way.

With any plausible tidal dissipation, very wide binaries generate a stellar collision once every $\sim$1000--7500 years in the Milky Way, potentially making them the dominant source of collisions in our Galaxy.  For comparison, evolving triple star systems yield one collision per $10^4$ yrs in the Galaxy, while the Milky Way's globular clusters' combined collision rate is one per $\sim$$10^6$ yrs  \citep{perkrat12}.  Considering both very wide binaries and triple stars, it appears that nearly all stellar collisions occur in the Milky Way field, which until recently had been considered devoid of collisions. 

\subsection{Effect of Stellar Encounter Velocity}

In most of our simulations we keep the local stellar dispersion fixed when generating stellar passages in our simulations.  In reality, the dispersion increases nearer to the Galactic center and decreases further away, and the typical field star encounter velocity can actually vary by over an order of magnitude throughout the Milky Way disk \citep{ver13}.  However, the exact dependence of local stellar dispersion on galactocentric distance is very uncertain and model-dependent \citep{bras10}.  In Section 4.1 we will demonstrate that the rate of periastron change in a very wide binary is independent of the local stellar dispersion as long as stellar encounters remain in the impulsive regime.  

Here we compliment this with an additional set of numerical experiments showing that the rate of stellar collisions within very wide binaries is relatively insensitive to the local stellar dispersion and is instead mostly set by the local stellar density.  Modeling the formation of the Oort Cloud at different galactocentric distances, \citet{bras10} assume that the local stellar velocity dispersion scales as $(8 / R_{G})^{0.92}$, where $R_{G}$ is the galactocentric distance measured in kpc.  In our additional simulations, we employ this model, scaling the \citet{rick08} stellar encounter dispersions as a function of the galactocentric distance of our simulated binaries.  This scaling causes the local stellar dispersion to vary by over a factor of 4 between our simulated binaries closest to the Galactic center and our outermost simulated binaries.  

\begin{figure}
\centering
\includegraphics[scale=.44]{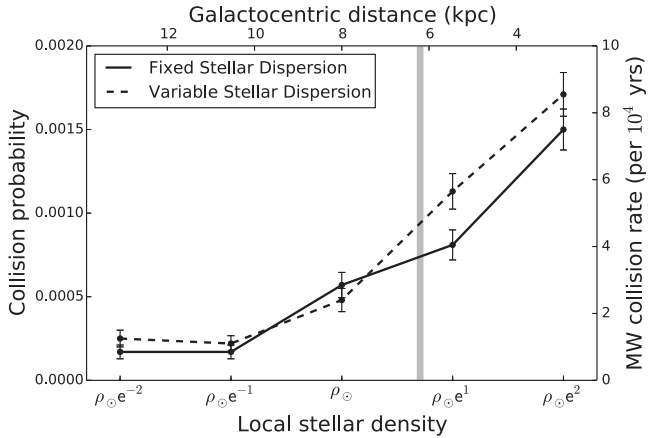}
\caption[S5]{The collision probability of simulated binary stars (with 1-$\sigma$ error bars) is plotted against their local stellar densities ($\rho_{\sun} \simeq 0.04$ M$_{\sun}$/pc$^3$).  Simulations using a fixed local stellar dispersion ({\it solid line}) are compared against those using a stellar dispersion dependent on galactocentric distance ({\it dashed line}).}
\label{fig:1}
\end{figure}

The collision rates seen in these new simulations are compared against our fiducial simulations in Figure 4 (both sets of simulations contain the same tidal dissipation prescription).  As can be seen, the collision rates in both sets of simulations are very similar as a function of galactocentric distance.  The simulations with a variable local stellar dispersion may have a slightly enhanced encounter rate in the interior of the Galaxy, but this variation is dwarfed by the sensitivity to tidal dissipation seen in Figure 2.  Thus, we can conclude that the stellar collision rates determined in our work are fairly independent of our assumptions of the typical field star encounter velocity.

\subsection{Formation of Compact Binaries via Tidal Circularization}

Figure 1B shows that instead of colliding, many binary companions pass close enough to each other that their orbital energy is rapidly dissipated by dynamic tides.  In this example, a binary's semimajor axis is shrunk from 8000 AU to less than 100 AU in just seven periastron passages.  Unlike the illustrative case displayed in Figure 1B, we normally stop following the evolution of binaries with $a<100$ AU for computing efficiency reasons.  In addition, at small $a$ our technique of applying an instantaneous tidal dissipation at periastron becomes questionable, and binaries also transition from the dynamic tidal regime to the equilibrium tidal regime at small $a$.  However, these binaries' semimajor axes will continue to shrink below what is modelled in our typical simulations.  This raises the prospect that some very wide binaries can eventually be transformed into very tight contact binaries.  

A simple estimate of the encounter distance necessary for this transformation can be made calculating the rate that dynamic tides can change a binary's semimajor axis.  In our idealized simulations, the energy dissipation due to dynamic tides, $\Delta E(q)$, is applied instantaneously at periastron passage.  The level of dissipation in our fiducial tidal model depends only on stellar mass, radius, and periastron and is independent of semimajor axis.  The energy dissipation rate is therefore given by the orbital period of the binary if the periastron remains constant.  This yields the following rate of change in semimajor axis:

\begin{equation}
\dot{a} = \frac{1}{\pi m_{1} m_{2}} \left(\frac{m_{1} + m_{2}}{G}\right)^{1/2}a^{1/2} \Delta E(q).
\end{equation}
From this expression, we can then calculate the time required for dynamic tides to transform a binary from an initial semimajor axis ($a_{i}$) to a final semimajor axis ($a_{f}$):

\begin{equation}
t_{c} = \frac{2\pi m_{1} m_{2}}{\Delta E(q)}\left(\frac{G}{m_{1} + m_{2}}\right)^{1/2}\left(a_{f}^{1/2} - a_{i}^{1/2}\right).
\end{equation}
If $a_{i}\gg a_{f}$ we can ignore the $a_{f}$ term on the righthand side of Equation 4 and obtain a simple estimate of the tidal circularization timescale.  

Assuming a ``typical'' very wide binary configuration of $m_{1} = m_{2} = 0.4$ M$_{\sun}$ and $a_{i} = 10^{4}$ AU, we calculate this circularization time as a function of binary periastron for our fiducial tidal dissipation model \citep{pt77,leeos86} in Figure 5.  This plot allows us to estimate the maximum contribution that very wide binaries can make to the population of compact binary systems in our Galaxy.  We see that for $q$ beyond 10 stellar radii the circularization timescale is greater than the age of the Galaxy.  This therefore requires periastron to be below $\sim$7 stellar radii for tidal circularization to be possible.  However, this estimate also assumes that periastron is fixed.  The minimal importance of $a_{f}$ in Equation 4 demonstrates that most of the circularization time is spent at large $a$.  Consequently, if $t_{c}$ is large there is plenty of time for field star encounters to alter the periastron during the tidal circularization process.  Sometimes these encounters could push the periastron inward, speeding up tidal circularization, but others  could pull the periastron beyond 7 stellar radii and freeze the process.  Figure 3 shows that no stars passing within $\sim$5 stellar radii of each other later go on to collide.  This suggests that for $q\lesssim5\bar{R_{*}}$ tidal circularization proceeds fast enough that periastron will remain nearly constant during this process.  We can therefore consider $q\simeq5\bar{R_{*}}$ to be the periastron threshold inside which tidal circularization efficiently takes place.

\begin{figure}
\centering
\includegraphics[scale=.45]{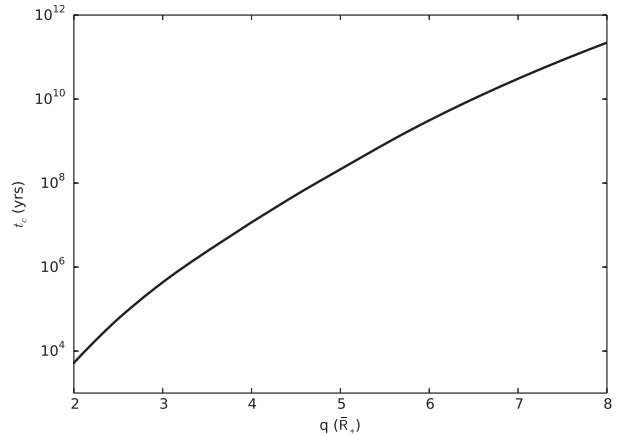}
\caption[S5]{Estimate of the tidal circularization timescale for a very wide binary with $m_{1} = m_{2} = 0.4$ M$_{\sun}$ and $a=10^4$ AU as a function of periastron for our fiducial dynamic tide model \citep{pt77,leeos86}.  Periastron is plotted in terms of the binary's mean stellar radius.  In this case, $\bar{R_{*}} = 0.4$ R$_{\sun}$.}
\label{fig:1}
\end{figure}

As mentioned previously, the intent of our simulations is to study star-star collisions, and they are not fully equipped to model the evolution of binaries down to very tight configurations.  However, we can estimate the maximum contribution using our numerical results.  To do this, we assume that every binary that tidally evolves to $a\le100$ in our simulations is destined to become a contact binary.  Based on this, we can estimate the maximum probability that very wide binaries evolve into contact binaries. 

We calculate these probabilities in Figure 6.  We see that the contact binary formation probability is of course dependent on the dynamic tidal dissipation we employ in our simulations as well as the local stellar density, which controls how often binaries are cycled into very eccentric orbits.  For our fiducial tidal model  \citep{pt77,leeos86}, we find that there is a median probability of 0.17\% that a very wide binary will evolve into a contact binary.  If we then optimistically assume that 10\% of stars reside in very wide binaries, we predict that 1 in $\sim$6000 stars will reside in contact binaries through our mechanism.  In contrast, modern surveys of contact binaries find that there is one contact binary for every 300--500 main sequence stars in the solar neighborhood  \citep{ruc02,get06}.  In our fiducial tidal model case, our mechanism falls about an order of magnitude short of explaining the contact binary fraction even with optimistic assumptions.  In fact, even our extreme tidal model, which almost certainly overestimates the degree of tidal circularization, cannot explain the observed contact binary fraction.  Although very wide binaries likely contribute to the formation of contact binaries, there must be more efficient formation mechanisms such as Kozai oscillations with tidal friction within triple star systems  \citep{kis98,eggkis01,fabtre07}.  Nevertheless, we find it very surprising that there exists a potential pathway for the very widest known binaries to evolve into the very tightest binaries.

\begin{figure}
\centering
\includegraphics[scale=.435]{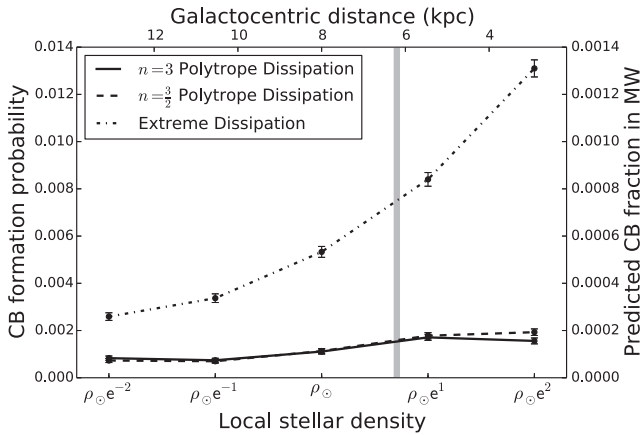}
\caption[S6]{The maximum contact binary formation vs the local stellar density for our simulated very wide binary systems.  Error bars correspond to 1-$\sigma$ Poisson uncertainties. The stellar density is written in terms of the Sun's local stellar density, $\rho_{\sun}$, which we assume to be 0.04 M$_{\sun}$/pc$^3$.  This is then converted to an approximate midplane galactocentric distance on the upper axis.  The grey shaded band marks the median local stellar density for thin disk stars assuming a disk scale length of 2.6 kpc.  Assuming a very wide binary fraction in the Milky Way of 10\%, we also plot the predicted contact binary fraction from this formation mechanism on the right axis.}
\label{fig:1}
\end{figure}

\subsubsection{Equilibrium Tidal Evolution}

As stated previously, we typically only consider the effect of dynamic tides  \citep{lai97, ivpap04} on the evolution of wide binaries.  We apply an instantaneous kick in energy at periastron passage in the relevant binaries (see Section 2.3).  This approximation is valid for very wide orbits with eccentricities close to unity, but breaks down as the stars approach each other.  In that situation, we must consider the regime of equilibrium tidal evolution.  Here we do so for a few illustrative systems.

Equilibrium tides account for the deformation of each of the binary stars by the other's gravitational potential.  The torques associated with tidal bulges, combined with energy dissipation within each star,
act to change the binary orbit  \citep{darwin79, hut81, egg98}.  Equilibrium tides act to shrink and circularize very eccentric binary orbits like the ones we consider.

The total angular momentum is conserved during tidal evolution. Assuming the orbital angular momentum $L$ to be the dominant component of the total angular momentum budget, a wide binary that is tidally interacting will be transformed into a close binary on a circular orbit with a semimajor axis of twice the periastron distance $q$. This is because $L \propto \sqrt{a (1-e^2)} = \sqrt{q(1+e)}$ and $e \approx 1$ for these binary orbits.

We calculate the late-stage evolution of a set of tidally-interacting wide binaries using an integrator employing an equilibrium tide model  \citep{bol12}.  The code uses the so-called ``constant time-lag'' model  \citep{hut81,egg98}.  The binaries are composed of equal-mass components with masses of 0.4 M$_\sun$ and corresponding radii of 0.412 R$_\sun$.  The stars are attributed dissipation values assuming pseudo-synchronous rotation with small obliquities  \citep{han10,bol12}, although we note that stellar dissipation rates are poorly-constrained.  We simulate nine binaries, each with an initial semimajor axis of 99 AU but initial periastron distances from 2.5 to 9 stellar radii.  As in Figure 1B, we assume that the transition from dynamic to equilibrium tides occurs at $a = $100 AU.  In reality, there will be a smoother transition between the two regimes.

Figure 7 shows the outcome of the integrations.  As expected, the orbits of the binaries are circularized at twice the initial periastron distance.  The timescale for the evolution depends sensitively on $q_0$.  Whereas the initially near-grazing binary with $q_0 = 2.5 R_{*}$ has its orbit circularized in $\sim$$10^5$ years, binaries with $q_0 > 6-7 R_{*}$ require hundreds of millions to billions of years for their orbits to be circularized.  Of course, this evolution is strongly dependent on the stellar parameters, in particular the dissipation rates within each star  \citep{hut81, egg98}.  Nonetheless, Figure 7 gives a reasonable order of magnitude estimate for the equilibrium tidal evolution timescales of these binaries.

\begin{figure}
\centering
\includegraphics[scale=.48]{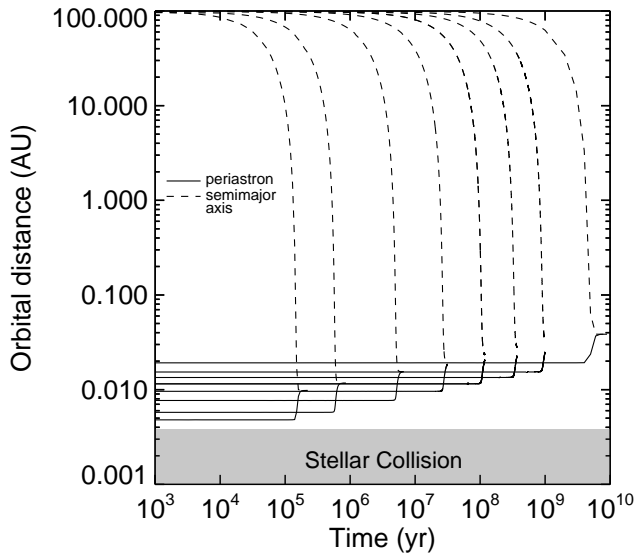}
\caption[S7]{Orbital evolution under the effect of equilibrium tides of a suite of binaries with eccentric orbits.  Each binary started the integration with a semimajor axis of 99 AU but with a different periastron distance $q_0$.  The faster-evolving binaries had smaller $q_0$.  From left to right, the curves correspond to $q_0$ = 2.5, 3, 4, 5, 6, 7, 8, and 9 times the sum of the stars' radii.  Integrations were stopped when the binary orbit was circular.}
\label{fig:1}
\end{figure}

\section{An Analytical Understanding}

\subsection{Dependence on Binary Semimajor Axis}

In our simulations, we model the orbital evolution of binaries with semimajor axes between 10$^3$ AU and $3\times10^4$ AU.  Here we calculate the collision probability of a very wide binary as a function of its semimajor axis.  Much of this derivation is nearly identical to the calculation of the flux of comets into the inner solar system from various regions of the Oort Cloud \citep{hills81}.  The reason these problems are so similar is because orbital perturbations from the local galactic environment drive both phenomena.  Further, instead of the Jupiter-Saturn barrier that prevents some eccentric comets from reaching the inner solar system, our binary systems have a dynamic tidal barrier whose dissipation prevents some eccentric binaries from ever reaching a collisional state.  

To begin, let's consider an ensemble of $N$ very wide binaries.  To simplify our problem, let's initially assume that these binaries all have the same age, stellar radius, and semimajor axis.  There are three ways to lose systems from this ensemble: collision, tidal circularization (see Figure 1B), and ionization from stellar impulses (which we will ignore for the moment).  We have seen from our simulations that very wide binaries are lost via tidal circularization if the stars passed within $\sim$5 stellar radii of each other in our fiducial model.  For a thermalized eccentricity distribution, the fraction of orbits in our binary ensemble with periastron less than 5 stellar radii is

\begin{equation}
F_{q} = \frac{10R_{*}}{a}
\end{equation}
where $a$ is the binary semimajor axis and $R_{*}$ is the typical stellar radius.  We will call this set of orbits the loss cone, since very wide binaries are lost through tidal circularization or potentially collisions if they occupy this orbital space.  Again if we assume that orbits are isotropized, then 2/5 of these binaries will actually collide rather than tidally circularize (since a periastron of two stellar radii or less yields a collision).  If binary orbits were distributed isotropically but did not evolve under external perturbations, this would be the total fraction of our binaries lost as a function of $a$.  However, these binary orbits are not static.  Consequently, new losses take place every orbital revolution as binaries with large initial periastron are perturbed to low periastron states.  

Like the Oort Cloud, the pericenter changes in our binaries are driven by perturbations from the Galactic tide and other passing field stars.  However, in our binary systems we are concerned with extremely tiny periastron values $\sim$3 orders of magnitude smaller than typical long-period comet pericenters.  In comet dynamics, the Galactic tide plays a major role in injecting comets into the inner solar system.  However, the periastron change per orbit due to the Galactic tide approaches zero as $q$ approaches zero \citep{lev06}.  Meanwhile, the tidal barrier requires relatively large shifts in periastron ($\Delta q\sim q$) for a collision to occur.  This renders the Galactic tide ineffective in producing stellar collisions.  Thus, we only need to consider the influence of impulses from passing field stars to estimate the fraction of binaries that collide.  

In the impulsive regime, the change in velocity a passing field star imparts on a binary star is approximately \citep{hills81} 

\begin{equation}
\Delta v = \left|\Delta v_{1} - \Delta v_{2}\right| = \frac{2Gm_{*}}{v_{*}}\left(\left|\frac{b_{1} - b_{2}}{b_{1}b_{2}}\right|\right)
\end{equation}
where $v$ is velocity, $b$ is the closest approach distance of the passing star, and $m_{*}$ is the passing star mass.  The subscripts 1, 2, and * refer to the primary, secondary and passing star respectively.  Averaged over all binary orbit orientations, $\left<\left|b_{1} - b_{2}\right|\right>\sim r$ and $\left<b_{1}b_{2}\right>\sim b_{1}^{2} \equiv b_{*}^{2}$.  Further, if $e\sim1$ then $\left<r\right>\sim1.5a$.  Therefore,

\begin{equation}
\left<\Delta v\right> \simeq 3Gam_{*} / v_{*}b_{*}^{2}
\end{equation}
Meanwhile, the orbital velocity of the binary companion at a distance $\left<r\right>$ is just 

\begin{equation}
v_{b} = \left(G\mu / 3a\right)^{1/2}
\end{equation}
where $\mu$ is the reduced mass of the binary.  

On average, a stellar encounter of a given impact parameter and velocity will change the orbital velocity of a binary by $\Delta v$.  Again, on average the magnitude of $\Delta v$ should be the same for all binaries if they have the same semimajor axes.  The direction of $\Delta v$, however, will depend on the orientation of the binary relative to the passing star.  Thus, the initial velocity vectors of an ensemble of randomly oriented binaries will be spread throughout a smear cone by a stellar passage (where the cone is centered upon the original velocity vector).  The angular size of this smear cone relative to the initial velocity direction is just

\begin{equation}
\theta = \frac{\left<\Delta v\right>}{v_{b}} = 9\frac{\mu v_{b}}{m_{*}v_{*}}\left(\frac{a}{b_{*}}\right)^{2}
\end{equation}
If $\Delta v\ll v_{b}$ then the fraction of all velocity space taken up by the smear cone is

\begin{equation}
F_{s} = \frac{\pi \theta^2}{4\pi} = \frac{27}{4}\left(\frac{m_{*}}{\mu}\right)^{2}\left(\frac{a}{b_{*}}\right)^{4}\left(\frac{G\mu}{av_{*}^{2}}\right)
\end{equation}

Meanwhile, for an isotropic velocity distribution we have already calculated the fraction of velocity space occupied by orbits leading to a collision or circularization in Equation 5.  Thus, we can now calculate the size ratio of the smear cone to the loss cone as

\begin{equation}
\frac{F_{s}}{F_{q}} = \frac{27}{40} \left(\frac{m_{*}}{\mu}\right)^{2} \left(\frac{a}{b_{*}}\right)^{4}\left(\frac{G\mu}{R_{*}v_{*}^{2}}\right)
\end{equation}
This ratio tells us how much of the loss cone is refilled with new binaries from a single stellar passage.  

If a loss cone is always filled ($F_{s}/F_{q}=1$) for an ensemble of binaries with a given $a$, then the fractional loss rate within this ensemble of binaries will just be set by the orbital period of the binaries and will be

\begin{equation}
\dot{L} = F_{q} \sqrt{\frac{G\left(m_{1} + m_{2}\right)}{4\pi^{2}a^{3}}}
\end{equation}
Thus, the loss rate (and therefore collision probability) decreases with increasing binary semimajor axis when the loss cone is always filled.  As the binary semimajor axis is decreased, the loss rate increases initially since the orbital period is shorter and $F_{q}$ is larger for tighter binaries.  However, tighter binaries also become less sensitive to field star perturbations, which are responsible for refilling the loss cone.  As a result, below some critical binary semimajor axis the loss cone will not always be filled, and this will act to decrease the loss rate and probability.  

We must therefore calculate how often stellar passages that are capable of refilling the loss cone for a given binary semimajor axis are expected to occur.  The required stellar passage strength can be calculated by setting $F_{s} / F_{q}$ to 1.  Thus, to refill the loss cone Equation 11 requires

\begin{equation}
\left(b_{*}^{2}v_{*}\right)^{2} \leq \frac{27}{40} \left(\frac{m_{*}}{\mu}\right)^{2}\left(\frac{G\mu a^{4}}{R_{*}}\right)
\end{equation}
Fortunately, the rate of stellar encounters with $b_{*}^2v_{*}$ less than a given quantity is very easy to calculate for a given stellar environment.  This rate is given by

\begin{equation}
f = \pi \left(b_{*}^{2}v_{*}\right) n_{*}
\end{equation}
where $n_{*}$ is the local spatial density of stars.  

As $f$ gets larger, the binary loss rate increases until $f$ exceeds the orbital frequency.  At this point the loss cone is ``saturated,'' and further increases in $f$ will not yield more collisions.  Thus, binaries with the highest probability of collision are those with semimajor axes $a_{crit}$ where $f$ exactly equals the orbital frequency.  Binaries with larger $a$ will have lower collision rates because of their longer orbital period and saturated loss cones.  Binaries with smaller $a$ will have lower collision rates because their loss cone is only occasionally filled.  

Setting $f$ equal to the orbital frequency, we can now solve for the stellar encounter strength that fills the loss cone once per orbital period for binaries with critical semimajor axis $a=a_{crit}$

\begin{equation}
\left(b_{*}^{2}v_{*}\right)^{2} = \frac{1}{4\pi^{4}n_{*}^{2}}\frac{G\left(m_{1}+m_{2}\right)}{a_{crit}^{3}}
\end{equation}
Now that we have re-expressed the stellar encounter strength in terms of the local stellar density and $a_{crit}$ we can substitute this expression into Equation 13 to obtain

\begin{equation}
a_{crit} = \left[\frac{10}{27\pi^{4}}\frac{R_{*}}{n_{*}^{2}}\frac{\mu\left(m_{1}+m_{2}\right)}{m_{*}^{2}}\right]^{1/7}
\end{equation}
Thus, the binary separation with the maximum loss (and therefore collision) probability decreases quite slowly with the local density of field stars ($a_{crit}\sim n_{*}^{-2/7}$).  Moreover, this critical separation does not depend at all on the velocity distribution of field stars as long as stellar encounters remain in the impulsive regime.  

Once $a_{crit}$ is determined it is easy to calculate the maximum loss probability that a binary can have for a given stellar environment and mass configuration.  This is just determined by the orbital period of $a_{crit}$, the size of the loss cone, and the age of the binary, $T_{age}$:

\begin{equation}
L = 1 - \left(1 - F_{q}\right)^{\frac{T_{age}}{2\pi}\sqrt{\frac{G\left(m_{1} + m_{2}\right)}{a_{crit}^{3}}}}
\end{equation}

We can also calculate how the binary loss probability changes for binary semimajor axes greater or less than $a_{crit}$.  When $a<a_{crit}$ the limiting factor is how often the loss cone is refilled.  In this regime loss cone refilling happens less than once per orbital period, so this portion of phase space will quickly empty and remain empty until another stellar encounter refills it.  Thus, 

\begin{equation}
L_{a<a_{crit}} = 1-\left(1-F_{q}\right)^{T_{age}f}
\end{equation}
where $f$ is given by Equations 13 and 14.  As long as $TfF_{q}\ll 1$ this probability can be approximated by
\begin{equation}
L_{a<a_{crit}} = T_{age}fF_{q}
\end{equation}
As mentioned above, 2/5 of the binaries that are lost through our loss cone will do so via collision.  Consequently, we only have to multiply Equation 19 by 2/5 to obtain a collision probability $P$ for binaries with $a<a_{crit}$:
\begin{equation}
P_{a<a_{crit}} = T_{age}m_{*}n_{*}a\sqrt{\frac{54 \pi^{2} G R_{*}}{5\mu}}
\end{equation}

Meanwhile, if $a>a_{crit}$ then the loss cone is always filled and the limiting factor is how often the binaries pass through periastron, i.e. the orbital period, $T_{orb}$.  In this case, Equation 17 yields the fraction of binaries that collide for a given $a$:

\begin{equation}
P_{a>a_{crit}} = \frac{2}{5}\left[1-\left(1 - F_{q}\right)^{T_{age} / T_{orb}}\right]
\end{equation}
Again, if $T_{age}F_{q} / T_{orb}\ll1$ then this probability is well-approximated by
\begin{equation}
P_{a>a_{crit}} = \frac{2}{5}\frac{T_{age}}{T_{orb}}F_{q} = \frac{2T_{age}}{\pi}\sqrt{\frac{G\left(m_{1}+m_{2}\right)R_{*}^{2}}{a^{5}}}
\end{equation}
Equations 20 and 22 are of course equal for $a=a_{crit}$.  

Thus, the collision probability of any binary can be estimated with Equations 16, 20, and 22.  These expressions can be further simplified if we assume all stars (including passing stars and our binary companions) to have some characteristic mass, $m_{*}$.  In this case, $m_{1} = m_{2} = 2\mu = m_{*}$.  With this assumption the expression for the binary semimajor axis with the highest collision probability becomes

\begin{equation}
a_{crit} = \left(\frac{10}{27\pi^{4}}\frac{R_{*}}{n_{*}^{2}}\right)^{1/7}
\end{equation}
while the probability of collision for any binary semimajor axis can be found with 

\begin{equation}
P_{a<a_{crit}} = T_{age}n_{*}a\sqrt{\frac{108 \pi^{2} G m_{*}R_{*}}{5}}
\end{equation}
and
\begin{equation}
P_{a>a_{crit}} = \frac{2T_{age}}{\pi}\sqrt{\frac{2Gm_{*}R_{*}^{2}}{a^{5}}}
\end{equation}

An example of this collision probability is plotted in Figure 8.  In this calculation, we set $m_{*}=0.4$ M$_{\sun}$ (the approximate mean field stellar mass), $R_{*}=0.4$ R$_{\sun}$, $T$ = 10 Gyrs, and $n_{*} = 0.1$ pc$^{-3}$ (the approximate spatial density of stars near the Sun).  We see in Figure 8 that the collision probability increases linearly with binary semimajor axis until $a=a_{crit}$, at which point the collision probability is roughly 1 in 300.  Beyond this, the collision probability falls off quickly with $a^{-5/2}$.  

\begin{figure}
\centering
\includegraphics[scale=.435]{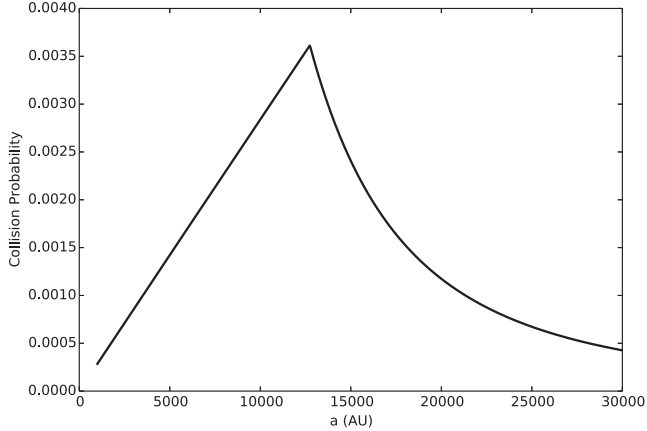}
\caption[S1]{The star-star collision probability as a function of binary semimajor axis predicted by Equations 23--25.  This particular distribution is generated assuming $m_{*}$ = 0.4 M$_{\sun}$, $R_{*} = 0.4$ R$_{\sun}$, $T_{age} = 10$ Gyrs, and $n_{*}$ = 0.1 pc$^{-3}$.  With these parameters  $a_{crit} = 12740$ AU, as indicated by the maximum collision probability.}
\label{fig:1}
\end{figure}

This probability distribution can then be integrated to yield a median semimajor axis for binary collisions as well as a total probability of binary collisions.  We have generated a distribution like the one shown in Figure 8 for each of the galactic environments we simulate.  Then we integrate these distributions over log-$a$ space (since we assume a log-uniform distribution of binary semimajor axes).  The median semimajor axis of colliding binaries is found where the normalized integral equals 0.5.  This predicted median semimajor axis is plotted as a function of local field star density in Figure 9.  In addition, we include the median binary semimajor axis found among our numerically simulated colliding systems in each environment.  As can be seen, the median semimajor axes found in both our numerical experiments and our theoretical predictions increase similarly with decreasing local stellar density.

\begin{figure}
\centering
\includegraphics[scale=.44]{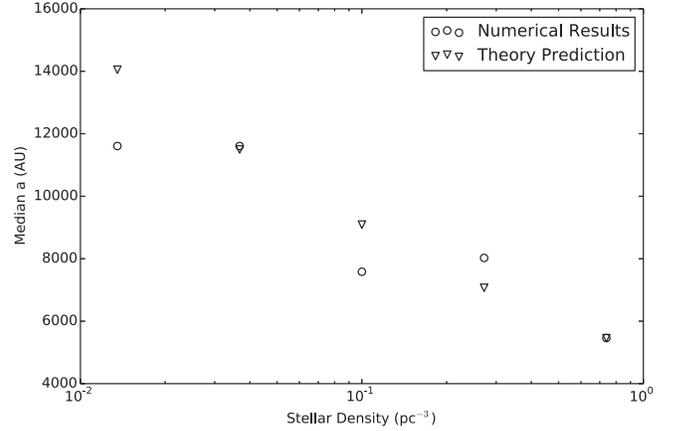}
\caption[S2]{The median binary semimajor axis of colliding binary companion as measured in our numerical simulations ({\it blue circles}) and as predicted by the probability distributions given in Equations 23--25 ({\it red triangles}).  The theoretical predictions are calculated assuming $m_{*}$ = 0.4 M$_{\sun}$, $R_{*} = 0.4$ R$_{\sun}$ and $T_{age} = 10$ Gyrs.}
\label{fig:2}
\end{figure}

In a similar fashion, we also integrate our probability distributions over all semimajor axes (in log-$a$ space) to obtain a predicted total collision probability for our binaries for each galactic environment.  These probabilities are plotted in Figure 10.  Here we see that our numerical results yield significantly lower collision probabilities than that predicted by Equations 23--25.  Moreover the disagreement becomes worse with denser environments.  A major reason for this discrepancy is that the above analysis neglected the ionization of binaries by field star impulses.  This acts to lower the collision probability because the binary lifetime can be substantially shortened via ionization, giving the binary a smaller time window to produce a collision.  Further, this effect is most pronounced in dense environments, where binaries are ionized more quickly.  

\begin{figure}
\centering
\includegraphics[scale=.44]{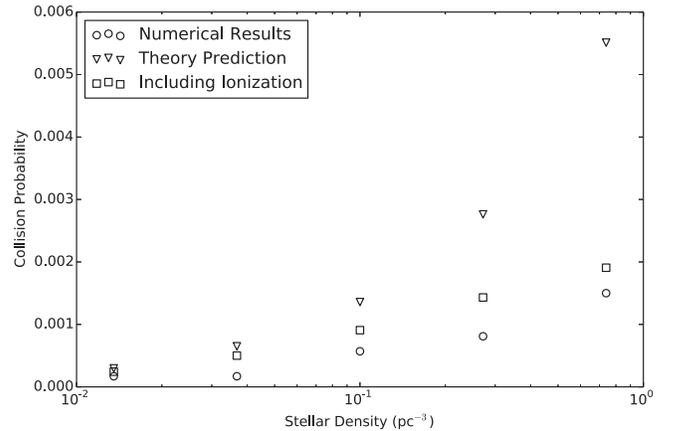}
\caption[S3]{The median binary semimajor axis of colliding binary companion as measured in our numerical simulations ({\it blue circles}), as predicted by the probability distributions shown in Equations 23--25 ({\it red triangles}), and as predicted by the probability distributions given in Equations 27 and 28, which account for the effects of binary ionization.  The theoretical predictions are calculated assuming $m_{*}$ = 0.4 M$_{\sun}$, $R_{*} = 0.4$ R$_{\sun}$, $T_{age} = 10$ Gyrs, and $v_{*} = 40$ km/s.}
\label{fig:3}
\end{figure}

We can attempt to account for this ionization by using the approximate formula for the ionization half-life of very wide binaries subject to stellar passages \citep{bah85}
\begin{equation}
t_{1/2} = 0.00233 \frac{v_{*}}{Gm_{*}n_{*}a}
\end{equation}
Assuming binary ionization is governed by exponential decay, Equations 24 and 25 can be modified to yield binary collision probabilities that account for the ionization of binaries.  These modifications yield the following new expressions for collision probability

\begin{equation}
P_{a<a_{crit}} = \tau n_{*}a\sqrt{\frac{108 \pi^{2} G m_{*}R_{*}}{5}}\left(1-{\rm e}^{-T_{age}/\tau}\right)
\end{equation}
and
\begin{equation}
P_{a>a_{crit}} = \frac{2\tau}{\pi}\sqrt{\frac{2Gm_{*}R_{*}^{2}}{a^{5}}}\left(1-{\rm e}^{-T_{age}/\tau}\right)
\end{equation}
where $\tau$ is the decay time constant ($t_{1/2} / \ln{2}$).  In Figure 10, we assume $v_{*} = 40$ km/s for the typical stellar encounter velocity and overplot the total collision probabilities predicted by our newly modified expressions.  There is a marked improvement in our predicted collision probabilities, especially for dense stellar environments.  The new predicted collision probabilities are now all within a factor of two of our numerically measured collision probabilities.  However, even this treatment does not fully account for the ionization process.  Since it is a slow diffusive process, binaries can diffuse out to larger $a$ over the course of our simulations without actually being ionized, and Figure 8 indicates that this diffusion to large $a$ will significantly decrease the collision probabilities of binaries that remain bound.  Nonetheless, we consider Figure 10 to be a very good match between simulations and theory given that all binary and passing stars are assumed to have the same fixed mass and that the stellar encounter velocity is also assumed to be fixed.  

\subsection{Galactic Tidal Precession vs. Equilibrium Tidal Precession}

As mentioned previously, the Milky Way's tide is ultimately ineffective in driving the periastron of  binaries to collision.  The final decrease in periastron during a collision is always caused by a stellar encounter.  However, the Galactic tide can still play a role in pushing binaries to the brink of collision before stellar encounters take over.  The direction of periastron change due to the Galactic tide is set by the value of the argument of pericenter, $\omega$, that the binary has with respect to the Milky Way midplane.  For most galactocentric distances (including the ones modeled here), the vertical disk term dominates over the radial tide component \citep{bras10}.  For binary eccentricities near 1 (as in the case of our colliding binaries), the precession rate of $\omega$ under the vertical component of the Milky Way's tide in a disk-only potential is approximately given by

\begin{equation}
\dot{\omega}_{GT} \simeq -\frac{5\sqrt{2G}\pi}{q\left(m_{1} + m_{2}\right)}\rho_{0}a^2 \sin^{2}{\omega}\cos^{2}{i}
\end{equation}
where $q$ is periastron and $i$ is inclination relative to the Galactic midplane \citep{heitre86,lev06}.

In addition to the Galactic tide, there are other sources of orbital precession.  These arise from the stars' tidal bulges as well as general relativistic effects.  If these sources of precession become comparable to that due to the Galactic tide, the dynamics of very wide binaries can diverge from that seen in our numerical simulations, since these additional precession sources are not included in our numerical work.  The precession of $\omega$ due to general relativity is given by the following expression \citep{fabtre07}

\begin{equation}
\dot{\omega}_{GR} = \frac{3G^{3/2}\left(m_{1} + m_{2}\right)^{3/2}}{a^{5/2}c^{2}\left(1-e^{2}\right)}
\end{equation}

Meanwhile, the precession due to the tidal bulges can be estimated from equilibrium tidal models to be \citep{fabtre07}

\begin{multline}
\dot{\omega}_{tide} = \frac{15\left[G\left(m_{1} + m_{2}\right)\right]^{1/2}}{8a^{13/2}}\frac{8+12e^{2}+e^{4}}{\left(1-e^{2}\right)^{5}} \times \\ \left(\frac{m_{2}}{m_{1}}k_{1}R^{5}_{1}+\frac{m_{1}}{m_{2}}k_{2}R^{5}_{2}\right)
\end{multline}
where $k$ is the Love number, $R$ is the stellar radius, and the subscripts 1 and 2 refer to the primary and secondary stars as usual.

To estimate the accuracy of our simulations we can now compare the magnitudes of the three precession rates given in Equations 29--31.  This is done in Figure 11.  To generate this plot we assume $m_{1}=m_{2}=0.4$ M$_{\sun}$, $R_{1}=R_{2}=0.4$ R$_{\sun}$, and $k_{1}=k_{2}=0.014$ (a typical estimate for stars \citep{fabtre07}).  In addition, we set $q=5R_{*} = 2$ R$_{\sun}$, since we already know from our numerical work that dynamic tides dominate over Galactic tidal effects inside of this periastron distance.  Thus, Figure 11 measures whether binary orbital precession is dominated by the Galactic tide right up until binary companions collide or tidally circularize.  

\begin{figure}
\centering
\includegraphics[scale=.67]{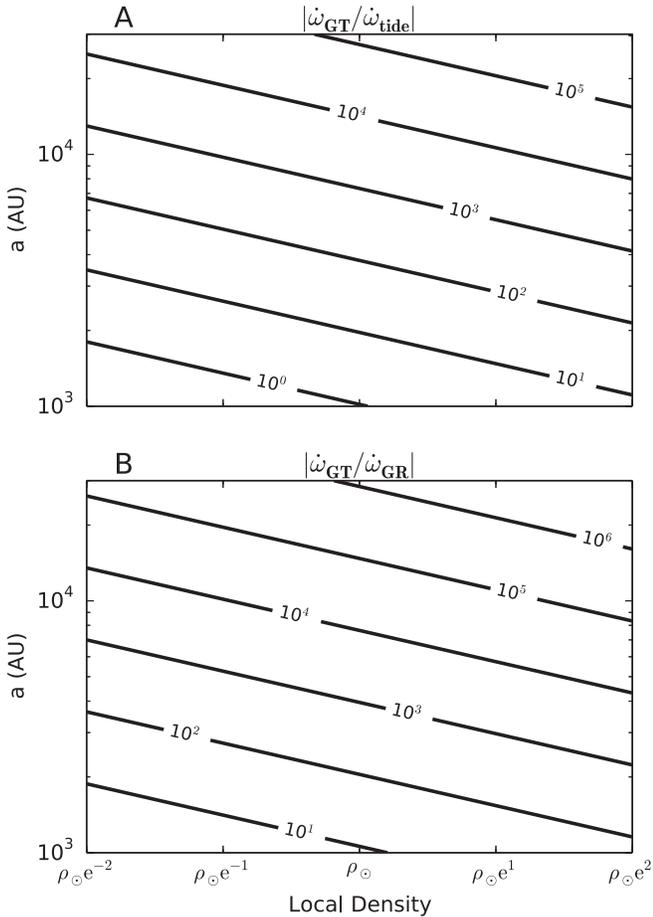}
\caption[S1]{{\bf A:} Contour plot of the ratio of the magnitude of the binary orbital precession rate due to the Galactic tide to the precession rate due to stellar tidal bulges as a function of binary semimajor axis and local galactic density.  {\bf B:} Contour plot of the ratio of the magnitude of the binary orbital precession rate due to the Galactic tide to the precession rate due to general relativity as a function of binary semimajor axis and local galactic density.  In these calculations, $\rho_{\sun}=0.04$ M$_{\sun}$/pc$^{3}$, $q = 5$ R$_{*} = 2$ R$_{\sun}$, $m_{1} = m_{2} = 0.4$ M$_{\sun}$, $R_{1} = R_{2} = 0.4$ R$_{\sun}$, and $k_{1} = k_{2} = 0.014$.}
\label{fig:1}
\end{figure}

Figure 11A demonstrates that precession due to the Galactic tide is at least an order of magnitude faster than that due to tidal bulges for almost any very wide binary orbit.  The exception is for binaries with $a\lesssim3000$ AU that are located in very low density environments.  Here the two precession rates are comparable when binary periastron is at 5 $R_{*}$ (near the dynamic tidal barrier).  Figures 8, 9, and 10 clearly show, however, that this region of parameter space accounts for only a very small fraction of the binary collisions seen in our simulations.  Similarly Figure 11B shows a nearly identical situation when comparing general relativistic precession with galactic tidal precession.  Only at very low galactic densities and $a$ near 10$^3$ AU are the two precession rates comparable.  Again, this is only considering binary orbits with $q=5$ $R_{*}$.  For more circular orbits, the Galactic tide will be much more dominant.  Thus, we can conclude that for the vast majority of the parameter space probed in our numerical simulations it is perfectly acceptable to ignore the orbital precession due to tidal bulges and relativity.  

\section{Potential Observational Signature}

Within clusters, collision products can stick out as blue stragglers \citep{benzhills87}, but the Milky Way field has no fixed stellar main sequence turnoff to identify these stars.  However, lithium abundance is another potential collision signature \citep{gleb10}.  Due to sheer numbers, low-mass stars ($\lesssim1$ M$_{\sun}$) are the most likely collision participants.  Unlike higher mass stars which destroy Li very inefficiently, low-mass stars quickly deplete their primordial Li, and a collision between two low-mass stars yields an anomalously Li-depleted high-mass star.  Further, there is also an observed correlation between Li abundance and stellar rotation rate, since both are stellar age proxies \citep{cut03}.  Most stellar collisions will be off-axis, forcing collision products to rotate extremely rapidly immediately after the collision \citep{lom96,sills01}.  (In fact, some may even form new circumstellar disks \citep{benzhills87,demarc04,sills05}, raising the possibility of a second generation of planet formation.)  In $\sim$20\% of our simulated collisions two stars with masses below 0.85 M$_{\sun}$ collide to yield a merged stellar mass above 1.2 M$_{\sun}$.  In Figure 2, our fiducial model predicts such collisions occur once every $\sim$12000 years, maintaining a small population of single, rapidly rotating, high-mass, Li-depleted stars that contradict the observed correlations between these stellar properties.  Meanwhile, conventionally formed high-mass stars typically have Li abundances 2--3 orders of magnitude higher than those with masses below 0.85 M$_{\sun}$ \citep{ram12}.  Our predicted population of collisionally formed high-mass stars should be 500--1000 times smaller than that formed via conventional star formation \citep{garc01}.  Instances of single, rapidly rotating, Li-depleted, high-mass stars have been discovered by stellar surveys \citep{cut03}.  At present there is no explanation for such stars, and the collision mechanism described in our work should be considered as a potential source. 

\section{Summary}

In this work, we have demonstrated that external perturbations from the Milky Way tide and passing field stars can often drive the stellar orbits of very wide binary stars through brief phases of very high orbital eccentricity.  In extreme instances of this, the periastron of binary systems can approach values comparable to or less than a stellar radius, and the two companion stars can physically collide.  Via numerical simulations of this process, we have found that a typical very wide binary star has a probability between 1 in 5000 and 1 in 500 of generating a stellar collision over the course of its life.  Assuming that very wide binaries represent 5\% of all stars in the Milky Way's thin disk, we predict that very wide binary companions collide once ever 1000--7500 years in our Galaxy, perhaps making these systems the dominant source of stellar collisions in the Milky Way.

The main impediment to collisions is dissipation due to dynamic tides when two companion stars first pass very near each other without colliding.  In these cases, the energy dissipated by tides can rapidly decrease the binary's semimajor axis, which effectively insulates the binary from additional galactic perturbations.  Since these perturbations are what was driving the evolution of the binary periastron, this new tighter binary will not yield a stellar collision.  The level of energy dissipation due to dynamic tides remains poorly understood.  However, when we employ a simple prescription used in many previous works we find that the collision rate of very wide binaries is decreased from once per 1000 years to once per $\sim$2500 years in the Milky Way.  Although this tidal model is uncertain, we find that it is very difficult for tidal dissipation to suppress collision rates to values below once per $\sim$7500 years.  This is because a significant fraction of wide binary companions never come within tens of radii of each other before the periastron passage that results in collision.  Suppressing such collisions would require a wildly unrealistic model of tidal dissipation.

In systems where collisions are avoided due to tidal dissipation, the very wide binaries very likely continue to become tighter and tighter due to continued tidal dissipation.  Since periastron is approximately locked in this process, these binaries are likely destined to eventually evolve to close or contact binaries.  Thus, surprisingly, there is a dynamical pathway to transform the widest binaries into the tightest ones.  However, our fiducial tidal model predicts that only as many as 10\% of contact binaries could be formed through this channel.  Thus, most contact binaries must be formed through another channel \citep{kis98,eggkis01,fabtre07}.  

Although there is no fixed main sequence in the Milky Way field, some of stellar collisions will give rise to very peculiar stars.  Because the collision is likely to be off-axis, the merged star must be rotating very rapidly initially.  In addition, low-mass stars have very low Li abundances, and these are our most likely collision progenitors.  Their merged  post-collision product will sometimes be a high-mass star.  These collisional products will contrast with normal high-mass stars, which are relatively rich in Li.  Thus, we predict a small population of Li-rich, rapidly rotating single stars will result from the collisions within very wide binaries.  We find it intriguing that stellar abundance surveys have uncovered potential evidence of such a population \citep{cut03}.

\section{Acknowledgements}
This work was funded by NSF Grant AST-1109776.  We thank Emeline Bolmont for her equilibrium tides code as well as the anonymous reviewer for insightful comments.  SNR thanks the CNRS's PNP program and the NASA Astrobiology Institute's Virtual Planetary Laboratory team.

\bibliographystyle{apj}

\bibliography{WBCollide}

\end{document}